\documentclass[12pt]{article}
\usepackage{graphicx}
\usepackage{epstopdf, epsfig}
\pdfoutput=1

\usepackage[a4paper, margin=2.5cm]{geometry}
\usepackage{amsmath} % For equation numbering
\usepackage{multirow} % To combine multiple rows
\usepackage{amssymb}
\usepackage{natbib}
\usepackage{caption}
\usepackage{bm}
\usepackage{tabularx}

\usepackage{placeins} % For float barriers
\usepackage{hyperref}
\usepackage[mathscr]{euscript}
 \let\mathscr\relax
\usepackage[scr]{rsfso}

\usepackage{enumitem}

\usepackage{cleveref}
\crefformat{section}{\S#2#1#3}

\title{Viscous and Inviscid Reconnection of Vortex Rings on Logarithmic Lattices}

\author{\small Abhishek Harikrishnan, Adrien Lopez, Bérengère Dubrulle}

\date{\small 
Universit\'e Paris-Saclay, CEA, CNRS, SPEC, 91191, Gif-sur-Yvette, France}

\begin{document}

% ------------------- TITLE ------------------- %

\maketitle

% ------------------- ABSTRACT ------------------- %

\begin{abstract}

\noindent To address the possible occurrence of a Finite-Time Singularity (FTS) during the oblique reconnection of two vortex rings, Moffatt-Kimura (MK) (\textit{J. Fluid Mech.,} 2019a; \textit{J. Fluid Mech.,} 2019b) developed a simplified model based on the Biot-Savart law and claimed that the vorticity amplification $\omega_{\text{max}}/\omega_0$ becomes very large for vortex Reynolds number $Re_{\Gamma} \geq 4000$. However, with Direct Numerical Simulation (DNS), Yao and Hussain (\textit{J. Fluid Mech.,} 2020) were able to show that the vorticity amplification is in fact much smaller and increases slowly with $Re_{\Gamma}$. The suppression of vorticity was linked to two key factors - deformation of the vortex core during approach and formation of hairpin-like bridge structures. In this work, a recently developed numerical technique called \textit{log-lattice} (Campolina and Mailybaev, \textit{Nonlinearity}, 2021), where interacting Fourier modes are logarithmically sampled, is applied to the same oblique vortex ring interaction problem. It is shown that this technique is not only capable of capturing the two key physical processes overlooked by the MK model but also other quantitative and qualitative attributes generally seen with DNS, at a fraction of the computational cost. Furthermore, the sparsity of the Fourier modes allows us to probe very large $Re_{\Gamma} = 10^8$ until which the peak of the maximum norm of vorticity, while increasing with $Re_{\Gamma}$, remains finite, and a blowup is observed only for the inviscid case. 

\end{abstract}

% ------------------- KEYWORDS ------------------- %
% This should be removed at the end of writing the paper.
% Author will add this while submission
% They will be included in the typesetting process.

%\begin{keywords}
%	Atmospheric Boundary Layer, Stratification, Coherent Structures, Global intermittency
%\end{keywords}

% -------------------1. INTRODUCTION ------------------- %

\newpage

\section{Introduction}
\label{sec: Introduction}

Vortex reconnection is described by \citet{yao2020physical} as a ``fundamental topology-transforming dynamical event''. Studying its mechanism is important for predicting the behavior of trailing vortices of an aircraft, understanding the turbulence cascade and more importantly, the occurence of finite-time singularities (FTS) in Euler or Navier-Stokes equations \citep{yao2022vortex}. The starting configuration ranges from simple vortex rings and vortex tubes to complex knotted and linked vortices such as the Hopf link and the trefoil knot. Numerical simulations are usually carried out with either direct numerical simulation (DNS) or simplified models. 

With a Biot-Savart (B-S) model, \citet{kimura2017scaling} and references therein, found that anti-parallel vortex filaments stretch out as they approach closely forming a tent-like (or pyramid-like) structure before reconnection regardless of the initial configuration, thereby suggesting a universal route. To desingularize the B-S integral, a `cut-off' parameter is added to the denominator. This implies that the integral will produce spurious results for length scales smaller than the cut-off, such as at the time of reconnection. \citet{moffatt2019towards} argued that the evolution at reconnection time depends only on the curvature, the core radius and the separation distance. They used the B-S law to obtain analytical expressions for the rate of change of these variables, resulting in a nonlinear dynamical system, hereafter referred as the MK model. With this and subsequent work \citep{moffatt2019towards2}, they suggested the possible occurrence of `physical' singularity for both inviscid (Euler) flow and viscous flows when the vortex Reynolds number $Re_{\Gamma} = \Gamma/\nu \geq 4000$. Here, $\Gamma$ is the circulation strength and $\nu$ is the kinematic viscosity. In particular, they note that the vorticity amplification, i.e., the ratio of maximum vorticity at some time $t_c$ to the initial vorticity $\omega_{\text{max}}/\omega_{0}$, takes very large values with increase in $Re_{\Gamma}$. This is important as it was shown with the theorem of \citet{beale1984remarks} that if a FTS occurs at a critical time $t_c$, then the maximum norm of vorticity becomes unbounded, i.e., $\int_{0}^{t_c} ||\bm{\omega}(\bm{x}, t)||_{\infty} = \infty$. Employing a similar setup with $Re_{\Gamma}$ up to 4000, \citet{yao2020singularity}, hereafter YH, used DNS to show that the vorticity amplification is, in fact, much smaller than that reported by \citet{moffatt2019towards2}. In particular, YH attributed the suppression of vortex growth to flattening of the vortex cores as they approach closely and the braking effect of the bridges, both of which were ignored in the MK model. 

Early work with DNS at low $Re_{\Gamma}$, for instance \citet{kida1991collision}, were instrumental in establishing a clear understanding of the physical mechanism of reconnection. The process is usually divided into three phases. In the first phase called \textit{inviscid advection}, the vortex rings approach each other due to self-and mutual induction and they undergo strong vortex stretching. The cores also flatten resulting in a dipole structure. Next, during the \textit{bridging} phase, the antiparallel vortex lines annihilate each other resulting in the formation of bridge-or hairpin-like structures in a direction orthogonal to the interaction. While the bridge structures rapidly recede away, they still remain connected by remnant threads from incomplete reconnection, resulting in the \textit{threading} phase. High $Re_{\Gamma} \leq 40,000$ simulations with vortex tubes \citep{yao2020physical, yao2022vortex} have revealed more intricate details with increasing $Re_{\Gamma}$ such as the formation of the $k^{-5/3}$ spectrum as a result of successive reconnections and the core flattening becoming more pronounced. Since DNS may become prohibitively expensive for larger $Re_{\Gamma}$, it is useful to have a simplified model capable of preserving the important physical processes elucidated above while being computationally cheaper. 

In this paper, we use a projection of Navier-Stokes equations on a set of logarithmically spaced discrete Fourier modes to achieve very large Reynolds number, at moderate numerical cost. Such projection was invented by \citet{campolina2021fluid} and termed logarithmic-lattice (or log-lattice, in short). Although it may superficially look like a mere 3D generalization of well-known shell models of turbulence (see review by \citet{biferale2003shell}), the projection on log-lattice is a mathematically well defined procedure, allowing to preserve the main symmetries and conservation laws of the original equations, without the need of adjustable parameters. The interest of this technique for the problem of FTS has already been demonstrated by \citet{campolina2018chaotic} who showed that Euler equation on log-lattice develops a blow-up at finite time. The blow-up is characterized by a chaotic attractor, that spans a wide range of scales, out-of-reach of present DNS (see figure 2 of \citet{campolina2018chaotic}). This unique combination of using the equations of motion in their original form and the ability to span a wide range of scales with few modes makes log-lattice an attractive tool for studying vortex reconnections and its links to FTS. 

In this work, initial conditions similar to \citet{moffatt2019towards} and \citet{kida1991collision} are used to study some quantitative and qualitative aspects of oblique reconnection of vortex rings for $Re_{\Gamma} \rightarrow \infty$ with log-lattices. The qualitative studies require reconstruction of physical space from lattice variables which is currently an open question. Previous studies with shell models \citep{bohr1998dynamical, gurcan2017nested} have suggested employing the standard discrete Fourier transform to reconstruct the velocity field in physical space. This method described in subsection \ref{subsecsec: Visualization} is useful to verify the initial conditions (cf. subsection \ref{subsecsec: init_conditions}) that need to be described directly in Fourier space. Along with the qualitative studies, temporal behavior of global quantities such as enstrophy $\mathcal{{E}}$ and maximum norm of vorticity $||\bm{\omega}||_{\infty}$ with increasing $Re_{\Gamma}$ are discussed in section \ref{sec: results_discussion}.

\section{Numerical framework and initial conditions}
\label{sec: init_conditions_numerics}

\subsection{Log-lattice framework}
\label{subsecsec: log_lattice}

\begin{figure}
    \centering
    \includegraphics[width=\linewidth]{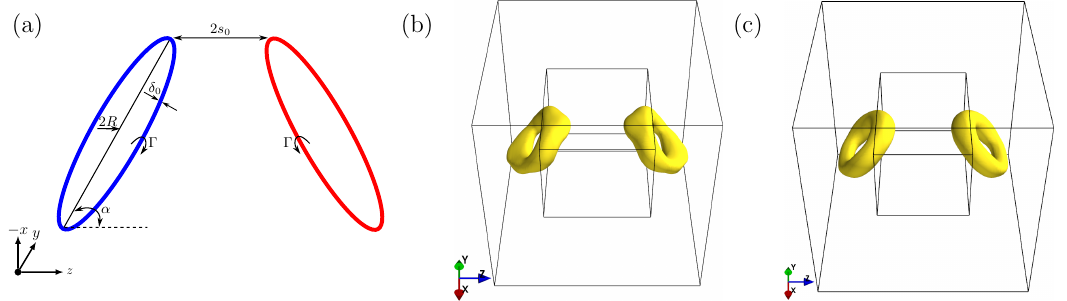}
    \captionsetup{width = \textwidth, justification=justified}
    \caption{(a) Schematic of the initial configuration. Figure adapted from YH. (b) Visualization of vortex rings from log-lattices for the Kida-type ring with $\delta_0/R = 0.2$ with the spacing factor  $\lambda = \phi$, (c) $\lambda \approx 1.194$ . Q-criterion isosurfaces on a grid of size $128^3$ are plotted at $\text{Q} > 0.35\, \text{Q}_{\text{max}}$ and $\text{Q} > 0.3\, \text{Q}_{\text{max}}$ respectively.}
    \label{fig:initConfig}
\end{figure}

Only a brief description of the log-lattice framework is given here. Additional details can be found in \citet{campolina2021fluid}. Starting with the Fourier transform of the incompressible Navier-Stokes equation,
\begin{align}
    \partial_t u_i + \bm{i} k_j u_i * u_j &= -\bm{i} k_i p - \nu k^2 u_i + f_i, \label{eq: Fourier_INSE_1}\\
    \bm{i} k_j u_j &= 0, \label{eq: Fourier_INSE_2}\\
    u_i * u_j (k) &= \sum_{\substack{q, r \in \Lambda^3 \\ q + r = k}} u_i(q) u_j(r) \label{eq: Fourier_INSE_3}
\end{align} 
where $\bm{i} = \sqrt{-1}$, $k_i$ is the $i$th component of the d-dimensional wave vector $\bm{k} = (k_1, ..., k_d)$, $p$ is the complex pressure field, $f_i$ denotes the forcing and $\nu$ is the kinematic viscosity. In this work, $f_i = 0$ for all simulations. When $\nu = 0$,  the flow is inviscid and the system reduces to the incompressible Euler equations. Here, (\ref{eq: Fourier_INSE_3}) describes the main convolution operation which couples Fourier modes in triadic interactions such that $k = q + r$ and $q, r, k$ are any three nodes on a \textit{logarithmic lattice} $\Lambda$. The logarithmic lattice is the set,
\begin{equation}
\Lambda = \{\pm \lambda^n\}_{n \in \mathbb{Z}}
\end{equation} \label{eq: log_lattice_set}
where $\lambda > 1$ is the spacing factor. As shown in \citet{campolina2021fluid}, nontrivial triad interactions exist only when the equation $\lambda^n = \lambda^q + \lambda^r$ has integer solutions for any $(n, q, r) \in \mathbb{Z}^3$. The following three families of solutions are known to exist, each with $z$ interactions in $d$-dimensions:

\begin{figure}
    \centering
    \includegraphics[width=0.9\linewidth]{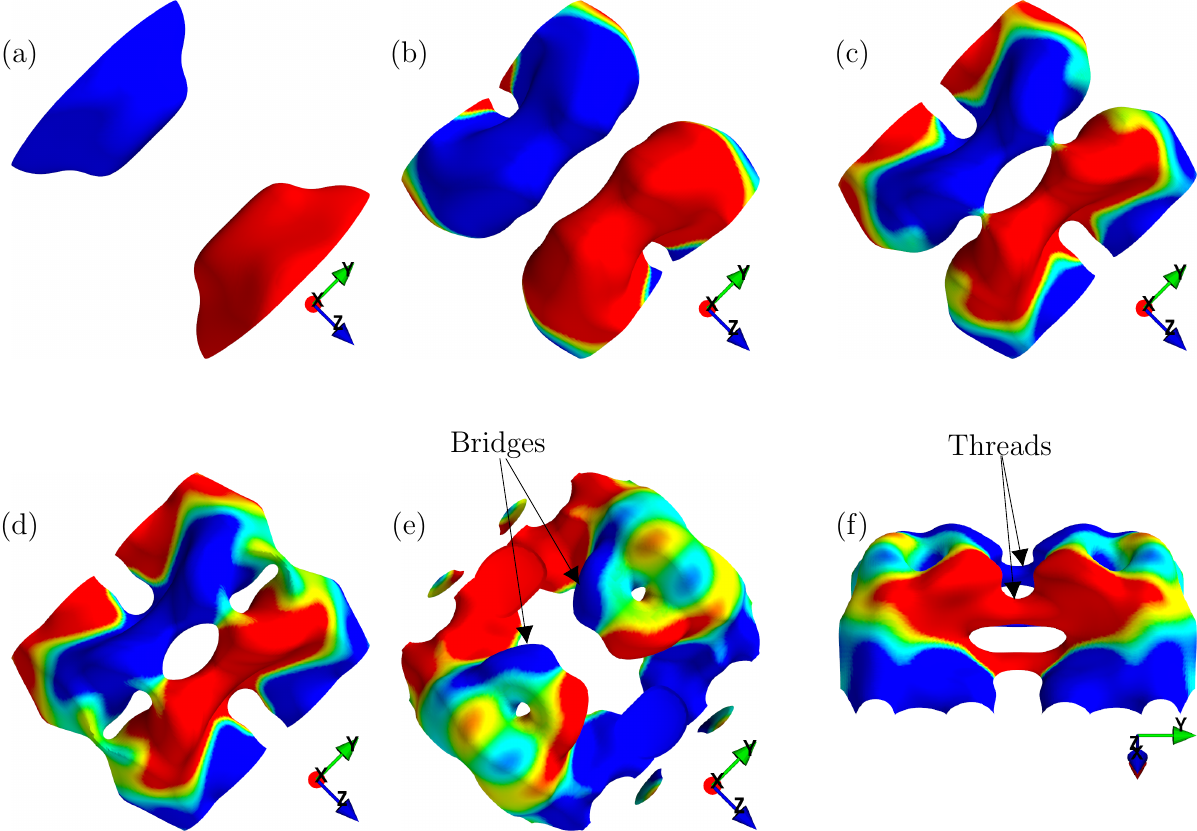}
    \captionsetup{width = \textwidth, justification=justified}
    \caption{Q-criterion isosurfaces shaded with contours of axial vorticity at $t = \{0.005, 0.118, 0.162, 0.175, 0.222, 0.201\}$ thresholded at $\text{Q} > 0.1\text{Q}_{\text{max}}$. Only a small portion of the domain is visualized as shown in figure \ref{fig:initConfig}(b).}
    \label{fig:Re1e4_qCrit}
\end{figure}

\begin{itemize}
    \item [(i)] $\lambda = 2$ with $z = 3^d$
    \item [(ii)] $\lambda = \sigma \approx 1.325$ is the plastic number with $z = 12^d$
    \item [(iii)] Any $\lambda$ that satisfies $1 = \lambda^b - \lambda^a$ where $(a, b)$ are mutual prime integers and $(a, b) \neq (1, 3), (4, 5)$. In this case, $z = 6^d$. For $(a, b) = (1, 2)$, the spacing factor $\lambda = \phi \approx 1.618$ is the golden mean. 
\end{itemize}

While decreasing the spacing factor $\lambda$ can increase the node density and the number of interactions per node which is desirable for simulating turbulent flows, this increases the computational cost. In this sense, simulations with $\lambda = 2$ are computationally cheaper but as pointed out in \citet{barral2023asymptotic}, it should be avoided for divergence-free flows because of the lack of backscatter. 

In this work, $\lambda = \phi$ is chosen for all simulations. The minimum wave vector $k_{\text{min}}$ is set to $2\pi$ to match a simulation on a box of size $L = 1$. The initial grid size is set as $(2N)^3 = 40^3$ for simulations without the zero component modes $\bm{k_0}$. The code is adaptive and the grid size is updated based on the fraction of energy contained in the outermost shells. The computational cost is reduced further by taking advantage of the grid symmetry along the initial axis $f(-k) = \overline{f(k)}$ where $\overline{(\cdot)}$ denotes complex conjugation. This means that the actual simulation is performed for $N \times 2N \times 2N$ nodes instead of $(2N)^3$ nodes for each velocity component. 

\subsection{Flow visualization with log-lattice}
\label{subsecsec: Visualization}

\begin{figure}
    \centering
    \includegraphics[width=0.975\linewidth]{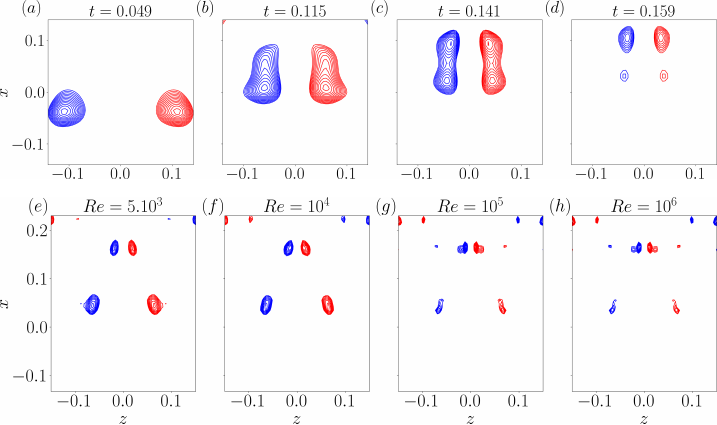}
    \captionsetup{width = \textwidth, justification=justified}
    \caption{(a-d) Temporal evolution of the vortex core shape is shown with contours of axial vorticity with level sets $\omega_y = [0.75-0.99]\omega_{y, \text{max}}$ in the symmetric (x, z) plane for the Kida-type ring at $Re_{\Gamma} = 10^4$. (e-h) Contours of axial vorticity at $t = 0.115$ and level sets $\omega_y = [0.85-0.99]\omega_{y, \text{max}}$ is shown in the symmetric (x, z) plane for the MK-type ring for increasing $Re_{\Gamma}$.}
    \label{fig:Re1e4_contour}
\end{figure}

To enable qualitative studies, 3D velocity fields need to be reconstructed in physical space from the lattice variables. As suggested in \citet{bohr1998dynamical, gurcan2017nested}, a simple algorithm would involve the discrete Fourier transform (DFT),
\begin{equation}\label{eq: inverse_DFT}
    \bm{u}(\bm{x}) = \sum_{\bm{k} \in \Lambda} {\bm{\hat{u}(k)}} . e^{i \bm{k}.\bm{x}} 
\end{equation} 
where $\bm{\hat{u}(k)}$ is the velocity vector in Fourier space, $\bm{k}$ is the non-uniformly spaced wave vector, $\bm{x}$ are evenly-spaced sample points. If $\Lambda$ is an evenly-spaced lattice, then (\ref{eq: inverse_DFT}) reduces to the regular DFT. While the complexity of the regular one-dimensional DFT is $\mathcal{O}(N^2)$, the complexity for log-lattices is $\mathcal{O}(N\log_{\lambda}N)$, similar to a Fast-Fourier transform, which reduces computation time. However, for large number of sample points, this is still computationally expensive and therefore, the python DFT code was GPU-parallelized with the CuPy library.

\subsection{Initial conditions and simulation details}
\label{subsecsec: init_conditions}

Two sets of initial conditions are chosen to study the oblique interaction of two vortex rings as shown in figure \ref{fig:initConfig}(a). The first set is similar to MK with initially circular vortex rings with radii $R$, thickness $\delta_0$ and inclination angle $\alpha = 45^{\circ}$. They have equal and opposite circulation $\Gamma$ and are separated at from their centerlines by a distance $2s_0 = 0.4$. The core size is $\delta_0/R = 0.01$. The only difference with the second set is the core size which is $\delta_0/R = 0.2$ which is similar to that studied by \citet{kida1991collision}. The larger core size enables better visualization of the various reconnection processes. 

Due to the spectral nature of log-lattices, the initial conditions need to be described in Fourier space as well. This immediately leads to the question of how to represent a vortex ring structure in Fourier space. An early idea exploited the fact that the Fourier transform of a 3D Gaussian function is another 3D Gaussian, and a rudimentary ring can be constructed by imposing the divergence-free condition. However, this type of ring has essentially one length scale to control its thickness but the radius of the ring cannot be controlled. The current idea makes use of the Dirac delta function to represent a 2D circle in 3D space. For a vortex ring of radius $R$, the vorticity is given by,
\begin{equation}\label{eq: vorticity_filament}
    \bm{\omega(x)} = \Gamma\int \delta^{(3)}(\bm{x} - \bm{R}(\phi)) d\bm{R}(\phi)d\phi
\end{equation}
where $\bm{R}(\phi) = (R\cos\phi, R\sin\phi, 0)$ and the unit tangent vector $d\bm{R}(\phi) = (-\sin\phi, \cos\phi, 0)$. Taking the Fourier transform of (\ref{eq: vorticity_filament}),
\begin{equation}\label{eq: vorticity_filament_FT}
    \bm{\hat{\omega}(k)} = \Gamma\int e^{-i \bm{k}.\bm{R}(\phi)} d\bm{R}(\phi)d\phi
\end{equation}
Using polar coordinates with $x = R\cos \phi$, $y = R\sin \phi$ and correspondingly $k_x = k_{\perp}\cos \alpha$, $k_y = k_{\perp}\sin \alpha$ such that $R = \sqrt{x^2 + y^2}$ and $k_{\perp} = \sqrt{k_x^2 + k_y^2}$, the dot product $\bm{k}.\bm{R}(\phi)$ reduces to $Rk_{\perp} \cos (\phi - \alpha)$ and (\ref{eq: vorticity_filament_FT}) can be written as,
\begin{equation}\label{eq: vorticity_filament_FT2}
    \bm{\hat{\omega}(k)} = \Gamma\int e^{-i Rk_{\perp} \cos(\phi - \alpha)} [-\sin \phi, \cos \phi, 0]d\phi
\end{equation}
Let $\theta = \phi - \alpha$ such that $d\theta = d\phi$ and after some trigonometric manipulation, we note that the result $\int e^{iRk_{\perp}\cos \theta} \cos \theta d\theta$ can be conveniently expressed as a Bessel function $-iJ_{1}(k_{\perp}R)$. Further, assuming $\Gamma = 1$ and convolving with a 3D Gaussian of width $\delta_0$ to represent the thickness of the vortex ring, the following closed-form expressions are obtained, 
\begin{align}\label{eq: vortex_ring}
    \hat{\omega}(k_x) &= -iJ_1(k_{\perp}R) \frac{k_y}{k_{\perp}} e^{\frac{-(||\bm{k}||\delta_0)^2}{2}} \\
    \hat{\omega}(k_y) &= iJ_1(k_{\perp}R) \frac{k_x}{k_{\perp}} e^{\frac{-(||\bm{k}||\delta_0)^2}{2}} \label{eq: vortex_ring_a}
\end{align}
This describes a single vortex ring in Fourier space having radius $R$ and thickness $\delta_0$ oriented along the z-direction. Straightforward extensions of the formula allow for translation and rotation of the ring. For instance, to translate the ring in the $-z$ direction by a distance $s_0$, one can convolve (\ref{eq: vortex_ring} - \ref{eq: vortex_ring_a}) with $e^{-ik_zs_0}$. Multiple rings can be added by superposition. Finally, the initial velocity field can be obtained by applying the Biot-Savart law,
\begin{equation}\label{eq:Biot-Savart}
    \bm{\hat{u}(k)} = \frac{i\bm{k} \times \bm{\hat{\omega}(k)}}{||\bm{k}||^2} \quad \text{for } \bm{k}\neq 0
\end{equation}
To study the qualitative aspects of reconnection, one needs to define a vortex. In this work, the second invariant of the velocity gradient tensor $\nabla \bm{u}$, i.e., the  Q-criterion \citep{hunt1988eddies} is chosen, which identifies vortices as regions where rotation dominates over the strain, 
\begin{equation}
    \text{Q} = \frac{1}{2} (||\bm{\Omega}||^2 - ||\bm{S}||^2) > 0
\end{equation}
where $\bm{\Omega} = \frac{1}{2} [\nabla \bm{u} - (\nabla \bm{u})^T]$ is the spin tensor and $\bm{S} = \frac{1}{2} [\nabla \bm{u} + (\nabla \bm{u})^T]$ is the strain-rate tensor. This can be seen as an immediate improvement over vorticity magnitude $||\bm{\omega}|| = ||\nabla \times \bm{u}||$, which is known to misidentify shearing motions as vortices as both regions possess non-zero vorticity \citep{lugt1979dilemma}. While numerous works in the vortex reconnection literature appear to favor the $\lambda_2$ criterion, which identifies regions as vortices when the middle eigenvalue $\lambda_2$ of the symmetric tensor $(\bm{S}^2 + \bm{\Omega}^2)$ possess negative values $\lambda_2 < 0$, \citet{chakraborty2005relationships} showed that it is possible to obtain equivalent thresholds among popular methods such as the Q, $\lambda_2$ and $\Delta$ criteria so that a particular choice among these criteria should not affect the qualitative results presented here.

The initial conditions for the Kida-type ring with $\delta_{0}/R = 0.2$, after applying the inverse DFT is visualized with the Q-criterion \ref{fig:initConfig}(b). Applying a threshold Q$>0.35$Q$_{\text{max}}$ reveals the vortex rings. One can take advantage of the fact that large scale modes are grouped closer on a logarithmic grid to ensure better physical space representation. This means using rings with large radii (for instance $R = 0.1L$ where $L$ is the integral scale) and large thickness lead to a better result. Further improvements can be made at the expense of increased computational cost, for instance, including the modes containing zero components $\bm{k_0}$ and reducing the grid spacing factor $\lambda$. All visualizations in this manuscript are shown with simulations performed with the $\bm{k_0}$ modes. The effect of reducing $\lambda$ can be seen from figure \ref{fig:initConfig}(c) where the rings appear more circular than \ref{fig:initConfig}(b). Due to the irregular spacing of the Fourier modes, using the inverse DFT also introduces artefacts/images with lower amplitude which are generally not visible at larger thresholds but become more apparent as the simulation progresses and turbulence decays. All visualizations and plots produced with real space data were made early times where the impact of these artefacts were negligible. To track the behavior of global quantities such as enstrophy and maximum norm of vorticity over long time, they were calculated directly in Fourier space which is unaffected by the artefacts.

\section{Results and discussion}
\label{sec: results_discussion}

\begin{figure}
    \centering
    \includegraphics[width=\linewidth]{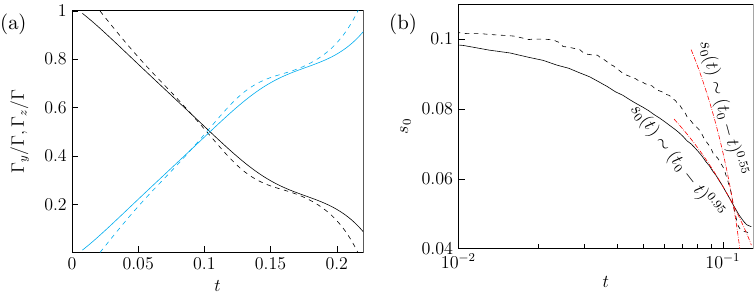}
    \captionsetup{width = \textwidth, justification=justified}
    \caption{Temporal evolution of (a) circulation $\Gamma_y, \Gamma_z$ normalized by the total circulation $\Gamma = \Gamma_y + \Gamma_z$ along the symmetric (x, z) plane shaded black and collision (x, y) plane shaded cyan respectively and (b) the separation distance $s_0$ is shown for both the Kida (solid line) and MK-type ring (dashed line) at $Re_{\Gamma} = 10^4$.}
    \label{fig:Re1e4_circ_separation}
\end{figure}

Figure \ref{fig:Re1e4_qCrit}(a-e) shows the temporal evolution of vortex reconnection within the smaller box indicated in figure \ref{fig:initConfig}b for the Kida-type ring at $Re_{\Gamma} = 10^4$. The Q-criterion fields are thresholded at $10\%$ of the maximum at each time step and shaded with axial vorticity $\omega_y$ to delineate the opposite rotation of the vortices. In the top panel, the first phase of reconnection, namely inviscid advection is evident. The anti-parallel vortex structures approach each other due to curvature-driven self-induction and collide. As explained in \citet{kida1991collision}, when the outermost vortex lines come into contact, they cancel each other at the point of contact due to viscous cross-diffusion. The process continues for other vortex lines forming hairpin-like bridge structures (see figure \ref{fig:initConfig}e) in a direction orthogonal to the initial approach of the two vortices. Since the hairpin-bridge structures are strongly curved at the tip, it generates a large self-induced velocity which pushes the structures outwards of the plane of the paper and backwards away from each other. This retreat also stops the cancellation of vortex lines and the remnant thread structures remain connected. Only the tail portion of the thread structure is visible as shown in figure \ref{fig:initConfig}f. 

The entire process including core compression is better illustrated in figure \ref{fig:Re1e4_contour}(a-d) where contours of axial vorticity along the symmetric plane (x, z) are plotted at various times. The core becomes flatter upon approach, resulting in a head-tail dipole structure. Extending the results of YH and contradicting MK, the vortex core flattens further and almost smears out with increasing $Re_{\Gamma}$, even when the core is thin, as evidenced in figure \ref{fig:Re1e4_contour}(e-h) where contours of axial vorticity are plotted for the MK-type ring at $t = 0.115$. 

\begin{figure}
    \centering
    \includegraphics[width=1\linewidth]{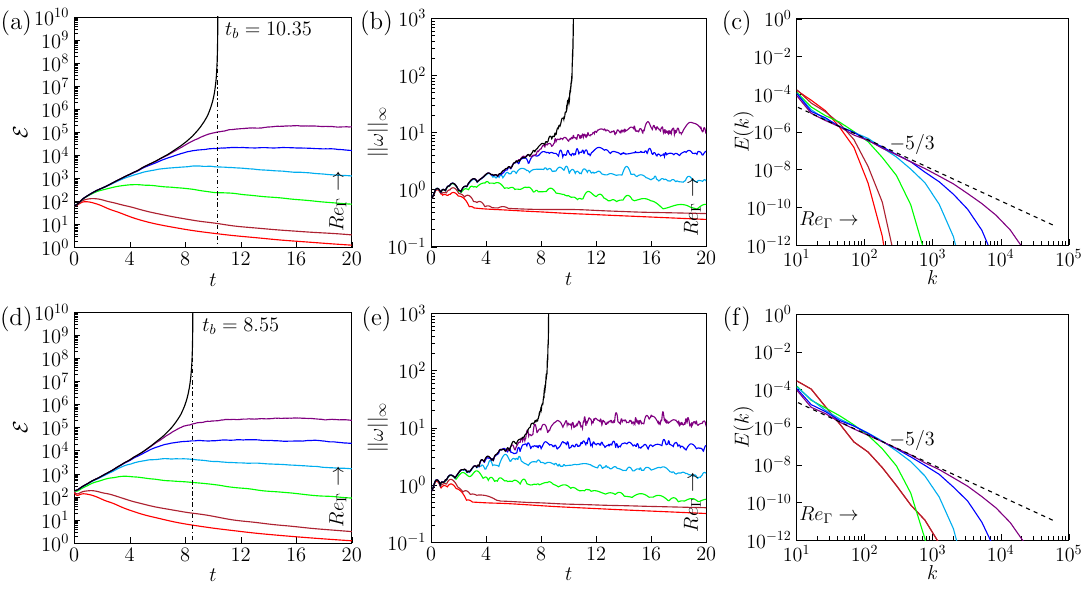}
    \captionsetup{width = \textwidth, justification=justified}
    \caption{Time evolution of (a, d) total enstrophy $\mathcal{E}$, (b, e) maximum norm of vorticity $||\omega||_{\infty}$ and (c, f) kinetic energy spectrum at peak enstrophy for increasing Reynolds numbers. The top and bottom panels correspond to the Kida and MK-type rings respectively. $Re_{\Gamma} = \{5.10^3, 10^4, 10^5, 10^6, 10^7, 10^8, \infty\}$ are shaded red, maroon, green, cyan, blue, violet and black respectively.}
    \label{fig:Kida_enstrophy}
\end{figure}

A key evidence of reconnection is the transfer of circulation from the symmetry (x, z) plane to the collision plane (x, y) plane. This is calculated for one half of each plane with $\Gamma_y = \int \bm{\omega}.\bm{n_s} dS$ and $\Gamma_z = \int \bm{\omega}.\bm{n_c} dS$, where $\bm{n_s}$ and $\bm{n_c}$ are unit vectors normal to the symmetric and collision planes, and plotted in figure \ref{fig:Re1e4_circ_separation}(a) for both Kida and MK-type rings at $Re_{\Gamma} = 10^4$. The inviscid advection phase is not evident as in YH due to the choice of the box (cf. figure \ref{fig:initConfig}(b)) where the complete vortex core is not visible during the initial time (see figure \ref{fig:Re1e4_qCrit}(a)). Therefore, the circulation along the symmetry plane initially increases as a larger portion of the vortex core comes into view. Later, a similar pattern as in YH is visible where the circulation continously drops along the symmetry plane while consequently increasing along the collision plane. 

An important question in reconnection studies is the rate of approach and separation of the vortices. If $\Gamma$ is the only relevant dimensional quantity, it was shown with dimensional analysis that the separation distance $s_0$ would scale as $s_0 \sim (\Gamma t)^{1/2}$. Indeed, for quantized vortex reconnections with the Gross-Pitaevskii model, the approach and separation rates were found to follow the same scaling suggesting that this scaling may be universal \citep{villois2017universal}. However, \citet{yao2020physical} note that for viscous vortex reconnection, this scaling depends on the core size with slender cores maintaining the local assumption required for the $1/2$ scaling. To examine the temporal evolution of separation distance during approach, an approach similar to YH is taken where the centroid of $\omega_y$ at $0.75\omega_{y, \text{max}}$ is calculated for one half of the symmetry plane and is taken to be the center of the tube. As shown in figure \ref{fig:Re1e4_circ_separation}(b), the separation distance before reconnection is found to scale $s_0 \sim (t_0 - t)^{0.95}$ for the Kida-type ring, far from the $1/2$ scaling obtained from dimensional analysis. However, as found in \citet{yao2020physical}, reducing the core size better maintains the local assumption and for the MK-type ring, $s_0$ scales as $(t_0 - t)^{0.55}$. 

The temporal evolution of enstrophy and maximum norm of vorticity are plotted in figure \ref{fig:Kida_enstrophy}(a, b) and figure \ref{fig:Kida_enstrophy}(d, e) for the Kida and MK-type rings respectively. As explained in subsection \ref{subsecsec: init_conditions}, these quantities are calculated directly from lattice variables in Fourier space. This means that they are free from artefacts allowing us to examine for long times. It is clear that both enstrophy and vorticity, while increasing with $Re_{\Gamma}$, remain finite even for very large $Re_{\Gamma}$. Only the Euler simulation (indicated with a solid black line) shows a blowup for both cases, albeit at different times. This difference is to be expected as it is already shown by \citet{pikeroen2024tracking} that the blow-up time is sensitive to initial conditions. Figures \ref{fig:Kida_enstrophy}(c, f) show the energy spectrum at peak enstrophy where one can observe a Kolmogorov-type $k^{-5/3}$ spectrum. At low $Re_{\Gamma}$, the range of the $-5/3$ slope is confined to the small wave numbers but this expands with increasing $Re_{\Gamma}$ suggesting a large number of small-scale structures being generated as a result of successive reconnections. 

\begin{table}
	\centering
    \setlength\tabcolsep{0pt}
	\begin{tabular}{|c | ccc | ccc|}
        \hline
		\multirow{3}{*}{$Re_{\Gamma}$} & \multicolumn{6}{c}{Vortex ring thickness} \vline \\
		\cline{2-7}
		& \multicolumn{3}{c}{$\delta_0/R = 0.2$} \vline  & \multicolumn{3}{c}{ $\delta_0/R = 0.01$} \vline \\
		\cline{1-7}
		& Initial grid size & Final grid size & CPU time (s)  & Initial grid size & Final grid size & CPU time (s)\\
        $10^4$ & $20\times 40\times 40$ & $20\times 40\times 40$ & $11.15$ & $20\times 40\times 40$ & $21\times 42\times 42$ & $18.27$ \\
        $10^8$ & $20\times 40\times 40$ & $29\times 58\times 58$ & $45.44$ & $20\times 40\times 40$ & $29\times 58\times 58$ & $21.31$ \\
        $\infty$ & $20\times 40\times 40$ & $64\times 128\times 128$ & $512.769$ & $20\times 40\times 40$ & $65\times 130\times 130$ & $428.87$ \\
        \cline{1-7}
        $\bm{10^4}$ & $\bm{21\times 41\times 41}$ & $\bm{21\times 41\times 41}$ & $\bm{11.23}$ & $\bm{21\times 41\times 41}$ & $\bm{22\times 43\times 43}$ & $\bm{17.42}$ \\
        $\bm{10^8}$ & $\bm{21\times 41\times 41}$ & $\bm{40\times 79\times 79}$ & $\bm{427.04}$ & $\bm{21\times 41\times 41}$ & $\bm{41\times 81\times 81}$ & $\bm{351.8}$ \\
        $\bm{\infty}$ & $\bm{21\times 41\times 41}$ & $\bm{68\times 135\times 135}$ & $\bm{1119.23}$ & $\bm{21\times 41\times 41}$ & $\bm{79\times 157\times 157}$ & $\bm{485.72}$ \\
        \hline
	\end{tabular}
    \captionsetup{width = \textwidth, justification=justified}
    \caption{Initial, final grid sizes and time taken for some cases of the log-lattice simulation. Bold font indicates simulations performed with the $\bm{k_0}$ mode. CPU time indicates the time taken for the convolution operation $(\ref{eq: Fourier_INSE_3})$ at the last time step of the simulation.}\label{tab: grid_time}
\end{table}

\section{Conclusion}
\label{sec: conclusion}

With the log-lattice technique, numerical simulations of two inclined vortex rings with core sizes $\delta_0/R = 0.2, 0.01$ are performed for increasing $Re_{\Gamma}$ up to $10^8$, along with an inviscid (Euler) simulation. It is capable of preserving key physical processes seen in DNS including core flattening and the formation of hairpin-like bridge structures that suppress vorticity amplification. In line with the results of YH, the peak of the maximum norm of vorticity increases with $Re_{\Gamma}$ but remains finite even at $Re_{\Gamma} = 10^8$ and a blowup is seen only for the inviscid case. Other qualitative results observed with DNS studies such as circulation transfer and separation distance scaling are also captured quite well by log-lattices making it a suitable simplified model to study vortex reconnections and its links to FTS at a much lower computational cost than DNS as shown in table \ref{tab: grid_time}. 

While it is possible to use the inverse DFT to visualize the reconnection process, this is limited to early times due to the presence of artefacts that are a result of using non-uniform, non-integer Fourier modes. Further improvements can be focused on interpolating the Fourier modes to linearly-spaced integer values to not only suppress artefacts but also to employ the standard Fast Fourier Transform algorithms. The delta function can also be used to study other initial conditions such as vortex tubes by replacing the equation of a circle with that of a line and convolving with a 3D Gaussian to give it some thickness. This is at the core of our current efforts to study the interaction of vortex tubes with increasing complexity - by varying the circulation strength, core sizes and introducing axial flow in the vortex cores at very high $Re_{\Gamma}$. Such studies of asymmetrical reconnection could be useful in devising a method for directly identifying vortex reconnection in turbulent flow data which is a major challenge.

\section*{Acknowledgements}

We thank Giorgio Krstulovic for suggesting the Dirac delta function to represent vortex rings in Fourier space.

\section*{Funding}

This research has been funded though the Agence Nationale pour la Recherche, via the grants ANR TILT grant agreement no. ANR-20-CE30-0035 and ANR BANG grant agreement no. ANR-22-CE30-0025.

\section*{Data and code availability statement}

The driver scripts and data generated from log-lattice simulations can be requested directly from the authors. The log-lattice python code used in the manuscript was developed by Amaury Barral as a part of his PhD work and is freely available along with a detailed documentation. See \citet{barral2024pyloggrid} for more details.

%% ------------------- BIBLIOGRAPHY ------------------- %
%
\bibliographystyle{unsrtnat} % Bibliography style
\bibliography{jfm.bib} % File where bibliography is stored

\end{document}